\RequirePackage{fix-cm}
\documentclass[twocolumn]{svjour3}          
\smartqed  
\usepackage{graphicx}
\usepackage{epstopdf}
\usepackage{subfigure}
\usepackage{amsmath}
\usepackage[misc]{ifsym}
\begin{document}

\title{An Improved Diversity Combining Receiver for Layered ACO-FOFDM in IM/DD Systems
}

\titlerunning{Photon Netw Commun}        

\author{Mengqi Guo \and Ji Zhou \and Xizi Tang \and Fan Hu \and Jia Qi \and Yaojun Qiao \and Aiying Yang \and Yueming Lu
}

\authorrunning{Photon Netw Commun} 

\institute{\Letter${\kern 1pt}$ Yaojun Qiao \at
${\kern 12pt}$qiao@bupt.edu.cn
\and
Mengqi Guo \and Ji Zhou \and Xizi Tang \and Fan Hu \and Jia Qi \and Yaojun Qiao \at
State Key Laboratory of Information Photonics and Optical Communications, School of Information and Communication Engineering, Beijing University of Posts and Telecommunications (BUPT), Beijing 100876, China\\
\email{guomengqi@bupt.edu.cn, zhouji@bupt.edu.cn, tangxizi@bupt.edu.cn, hufan30@bupt.edu.cn, qijia@bupt.edu.cn, qiao@bupt.edu.cn}
\and
Aiying Yang \at
School of Opto-electronics, Beijing Institute of Technology (BIT), Beijing 100081, China\\
\email{yangaiying@bit.edu.cn}
\and
Yueming Lu \at
Key Laboratory of Trustworthy Distributed Computing and Service, Ministry of Education, School of Information and Communication Engineering, Beijing University of Posts and Telecommunications (BUPT), Beijing 100876, China\\
\email{ymlu@bupt.edu.cn}
}

\date{Received: 11 March 2017 / Accepted: 8 July 2017}

\maketitle

\begin{abstract}
In this paper, an improved receiver based on diversity combining is proposed to improve the bit error rate (BER) performance of layered asymmetrically clipped optical fast orthogonal frequency division multiplexing (ACO-FOFDM) for intensity-modulated and direct-detected (IM/DD) optical transmission systems. Layered ACO-FOFDM can compensate the weakness of traditional ACO-FOFDM in low spectral efficiency, the utilization of discrete cosine transform in FOFDM system instead of fast Fourier transform in OFDM system can reduce the computational complexity without any influence on BER performance. The BER performances of layered ACO-FOFDM system with improved receiver based on diversity combining and DC-offset FOFDM (DCO-FOFDM) system with optimal DC-bias are compared at the same spectral efficiency. Simulation results show that under different optical bit energy to noise power ratios, layered ACO-FOFDM system with improved receiver has 2.86dB, 5.26dB and 5.72dB BER performance advantages at forward error correction limit over DCO-FOFDM system when the spectral efficiencies are 1 bit/s/Hz, 2 bits/s/Hz and 3 bits/s/Hz, respectively. Layered ACO-FOFDM system with improved receiver based on diversity combining is suitable for application in the adaptive IM/DD systems with zero DC-bias.

\end{abstract}

\section{Introduction}
Owing to the explosive growth of bandwidth hungry services, high capacity and low cost optical transmission systems are developed to meet the increasingly updated demands. As a multicarrier modulation format, orthogonal frequency division multiplexing (OFDM) can provide relatively high transmission capacity due to its superiorities in high spectral efficiency and robustness against chromatic dispersion and polarization-mode dispersion \cite{OFDM_Armstrong,OFDM_Shieh,OFDM_Djordjevic,A cost-effective and efficient}. The intensity-modulated and direct-detected (IM/DD) system has lower cost and power consumption compared with coherent system, it can be applied in many scenarios such as metropolitan area networks, access networks, data center interconnects and visible light communications \cite{metropolitan area network,PON,data center,VLC}. The signal transmitted in IM/DD OFDM system must be real and positive. To obtain real signal, Hermitian symmetry is needed to constrain the input constellations of inverse fast Fourier transform (IFFT). To obtain positive signal, DC-offset OFDM (DCO-OFDM) and asymmetrically clipped optical OFDM (ACO-OFDM) systems are most commonly used two schemes to generate positive signal \cite{DC-biased,ACO}.

The performance of DCO-OFDM depends on the DC-bias level. If the DC-bias is not large enough, the remaining negative values are clipped at zero level, which can introduce clipping distortions. If the DC-bias is quite large, this large DC-bias inefficiently occupies lots of optical power. The optimal DC-bias depends on the constellation size, which limits the performance of DCO-OFDM in adaptive systems. ACO-OFDM needs zero DC-bias, all the negative values are clipped at zero level in spite of the constellation size, so that ACO-OFDM is more suitable in adaptive systems. However, in ACO-OFDM system, only odd subcarriers are used to carry the signal, which leads to the loss of spectral efficiency.

To improve the spectral efficiency of ACO-OFDM, different types of multilayered schemes based on ACO-OFDM have been proposed. Asymmetrically clipped DC biased optical OFDM (ADO-OFDM) system transmits ACO-OFDM on odd subcarriers and DCO-OFDM on even subcarriers simultaneously, but the even subcarriers still need DC-bias \cite{Comparison of ACO-OFDM}. Hybrid ACO-OFDM system utilizes ACO-OFDM on odd subcarriers and pulse-amplitude-modulation discrete-multi-tone (PAM-DMT) on even subcarriers, but the real components of even subcarriers are still useless \cite{Hybrid asymmetrically clipped}. The superposition of more than two layers are proposed further to increase the spectral efficiency. Layered ACO-OFDM system applies different kinds of ACO-OFDM to different layers \cite{Layered ACO-OFDM}. Augmented spectral efficiency discrete multitone (ASE-DMT) system uses PAM-DMT on imaginary components of all the subcarriers, and employs layered ACO-OFDM on the real components \cite{Augmenting}.

It is worth noting that the performance of ACO-OFDM signal can be improved by extracting useful information from clipping noise on even subcarriers. A novel technique called diversity combining has been proposed in \cite{Diversity combining}, it utilizes the information from both of the odd and even subcarriers, which can also be used in multilayered schemes based on ACO-OFDM to further reduce the effect of noise and achieve better bit error rate (BER) performance.

In recent years, fast OFDM (FOFDM) based on discrete cosine transform (DCT) has been investigated in the IM/DD optical communication systems to reduce the computational complexity \cite{fast-OFDM1,fast-OFDM2,Asymmetrically Clipped Optical Fast OFDM,FOFDM}. If the inputs of DCT or inverse DCT (IDCT) are real values, the outputs are also real values. Therefore, the real transformation does not need Hermitian symmetry any more, and the one-dimensional modulation has lower computational complexity. For FOFDM system, the interval between subcarriers decreases to half of that in OFDM system, but signal on the positive frequency of FOFDM system has corresponding image on negative frequency, so the M-PAM FOFDM system has the same spectral efficiency and BER performance as M$^2$-QAM OFDM system \cite{Asymmetrically Clipped Optical Fast OFDM,FOFDM}. Consequently, it is not difficult to find DCT is also very suitable for replacing FFT in multilayered schemes to reduce the computational complexity without any influence on BER performance \cite{ZTE}.

In this paper, we firstly propose an improved receiver based on diversity combining technique in layered asymmetrically clipped optical FOFDM (ACO-FOFDM) for IM/DD optical transmission systems. Layered ACO-FOFDM system improves the spectral efficiency of traditional ACO-FOFDM system, and the utilization of DCT in FOFDM system instead of FFT in OFDM system can reduce the computational complexity without any influence on BER performance. Simulation results show that the layered ACO-FOFDM system with improved receiver not only has better BER performance than the system without improved receiver, but also has better BER performance than DCO-OFDM system with the optimal DC-bias and the same spectral efficiency.

\section{Transmitter Structure of Layered ACO-FOFDM}
\begin{figure*}[!t]
\centering
\includegraphics[width=14cm]{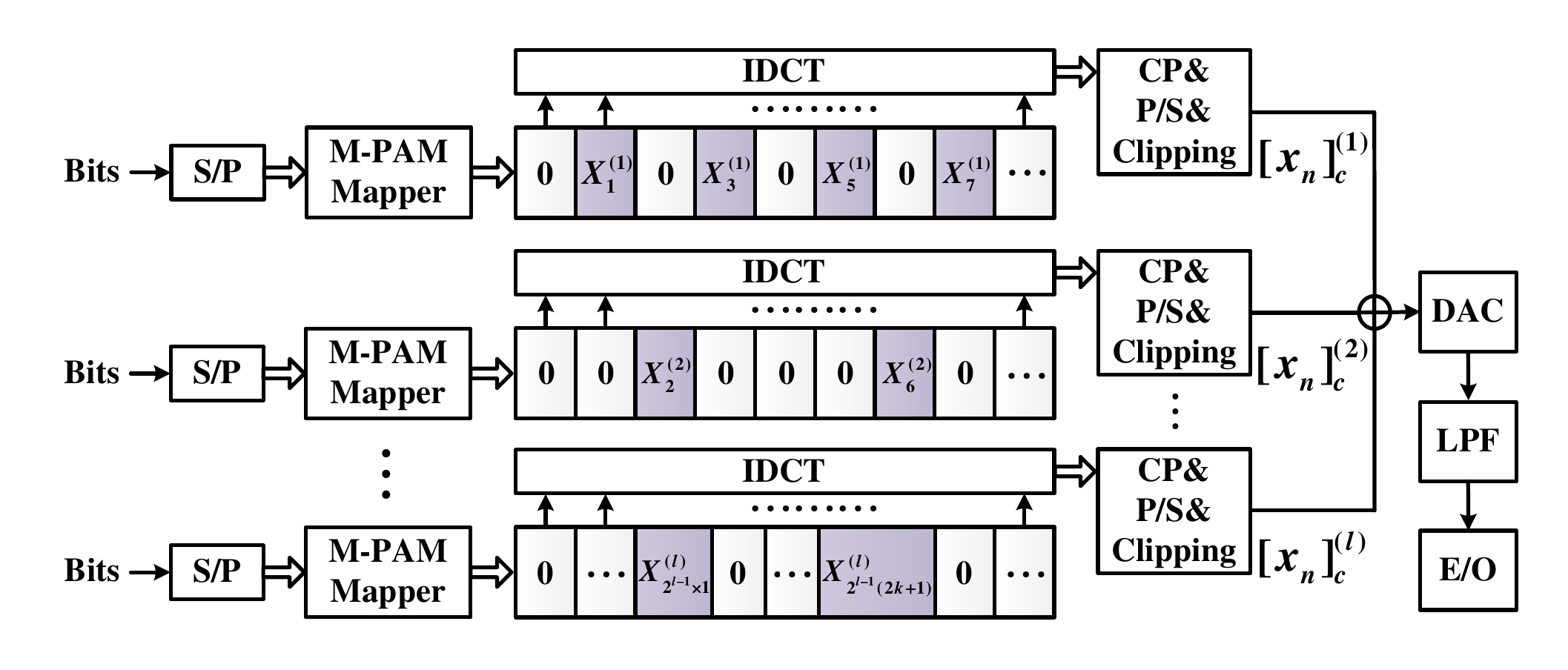}
\caption{Layered ACO-FOFDM transmitter scheme. $X_k^{(l)}$ denotes frequency domain signal on the $k$-th subcarrier of layer $l$, $[{x_n}]_c^l$ denotes clipped time domain signal on the $n$-th subcarrier of layer $l$}
\label{transmitter}
\end{figure*}

\begin{figure*}[!t]
\centering
\includegraphics[width=14cm]{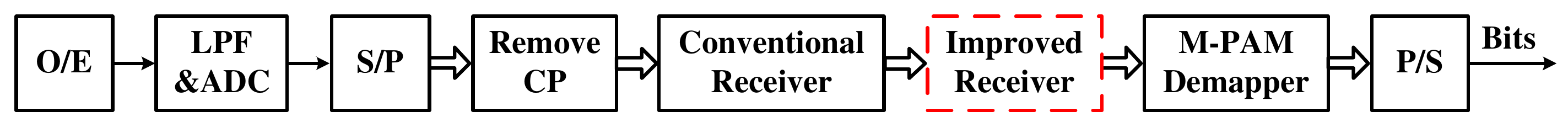}
\caption{Layered ACO-FOFDM receiver scheme}
\label{receiver}
\end{figure*}

For traditional ACO-FOFDM system, only half of the subcarriers are used to carry the signal, which leads to the spectral efficiency of ACO-FOFDM is half of that of DCO-FOFDM with the same modulation format. In order to improve the spectral efficiency of ACO-FOFDM, layered ACO-FOFDM system is proposed, so that signal on different layers can be transmitted simultaneously. Fig. \ref{transmitter} reveals the transmitter scheme of layered ACO-FOFDM system. The uppercase $L$ represents the total number of layers and the lowercase $l$ represents the $l$-th layer, these notations are employed throughout this paper.

The superposition of multiple layers is performed in frequency domain. Signal on the $1$-st layer is the same as that on the traditional ACO-FOFDM, the $l+1$-th layer occupies half number of subcarriers as the $l$-th layer does. As depicted in Fig. \ref{transmitter}, for layer 1, the original bits perform serial to parallel conversion and M-PAM mapping at first, then only odd subcarriers are utilized to carry the signal (index $2k+1, k=0,1,2,...,N/{2^1}-1$). The real-valued time domain signal is generated after IDCT, cyclic prefix is added and parallel to serial conversion is performed. Then, all the negative value of this odd-symmetry time domain signal is clipped at zero level. The clipping noise only falls on the even subcarriers (index $2k, k=0,1,2,...,N/{2^1}-1$), which is incapable to influence the signal on odd subcarriers. For layer 2, the odd subcarriers of remaining unused subcarriers on layer 1 are used (index $2\times(2k+1), k=0,1,2,...,N/{2^2}-1$), and the clipping noise falls on the even subcarriers of remaining unused subcarriers on layer 1 (index $2\times2k, k=0,1,2,...,N/{2^2}-1$). According to this rule, for layer $l$, only the ${2^{l-1}}\times(2k+1){\kern 2pt}(k=0,1,2,\ldots,N/{2^l}-1)$ subcarries are utilized to carry the signal, the clipping noise falls on the ${2^{l-1}}\times(2k){\kern 2pt}(k=0,1,2,\ldots,N/{2^l}-1)$ subcarries. Finally, unipolar time domain signals on different layers are generated, signals and clipping noise of the $l$-th layer are orthogonal to those on layers $1$ to $l-1$.

Afterwards, the unipolar time domain signals on different layers are added to transmit simultaneously. After digital-to-analog conversion (DAC) and low pass filter (LPF), the analog electrical signal is modulated to optical carrier.

\section{Receiver Structure of Layered ACO-FOFDM}

Figure \ref{receiver} reveals the receiver scheme of layered ACO-FOFDM system. The traditional layered ACO-FOFDM system only needs conventional receiver before M-PAM demapper. To improve the BER performance of layered ACO-FOFDM system, improved receiver is appended to conventional receiver.

\subsection{Conventional Receiver}
\begin{figure}[!t]
\centering
\subfigure[]{\includegraphics[width=\linewidth]{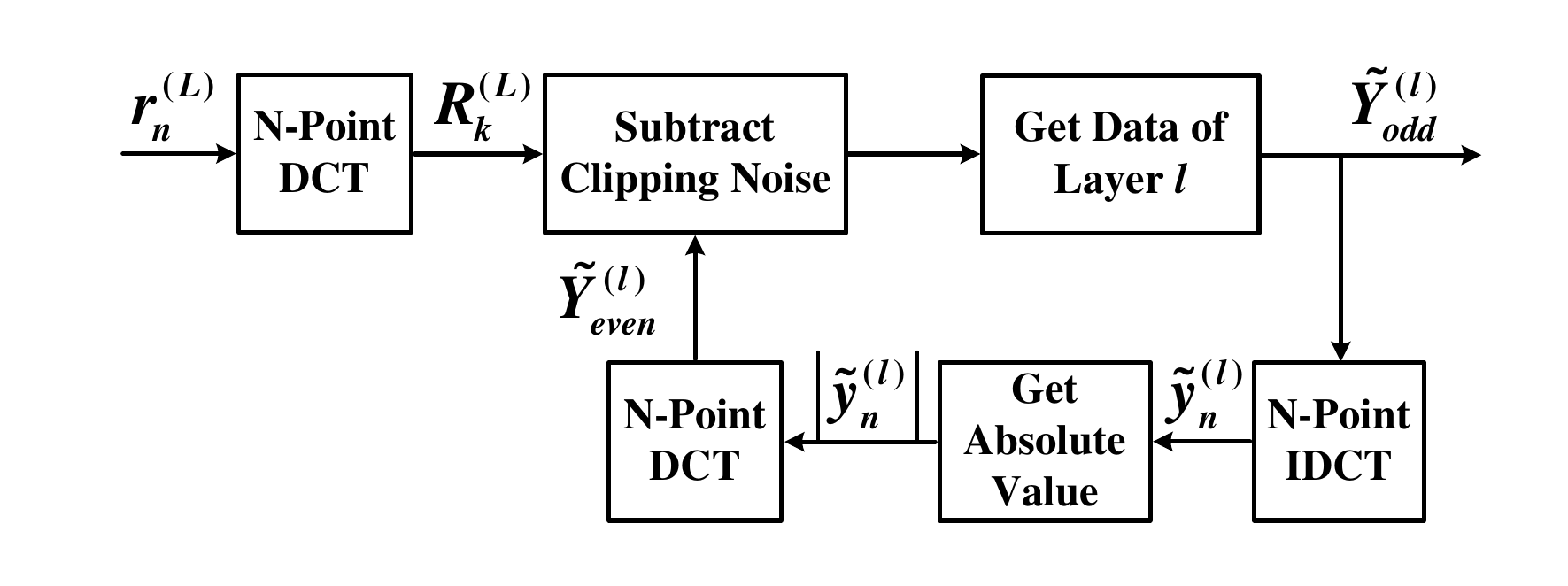}
\label{conventional-receiver}}
\subfigure[]{\includegraphics[width=\linewidth]{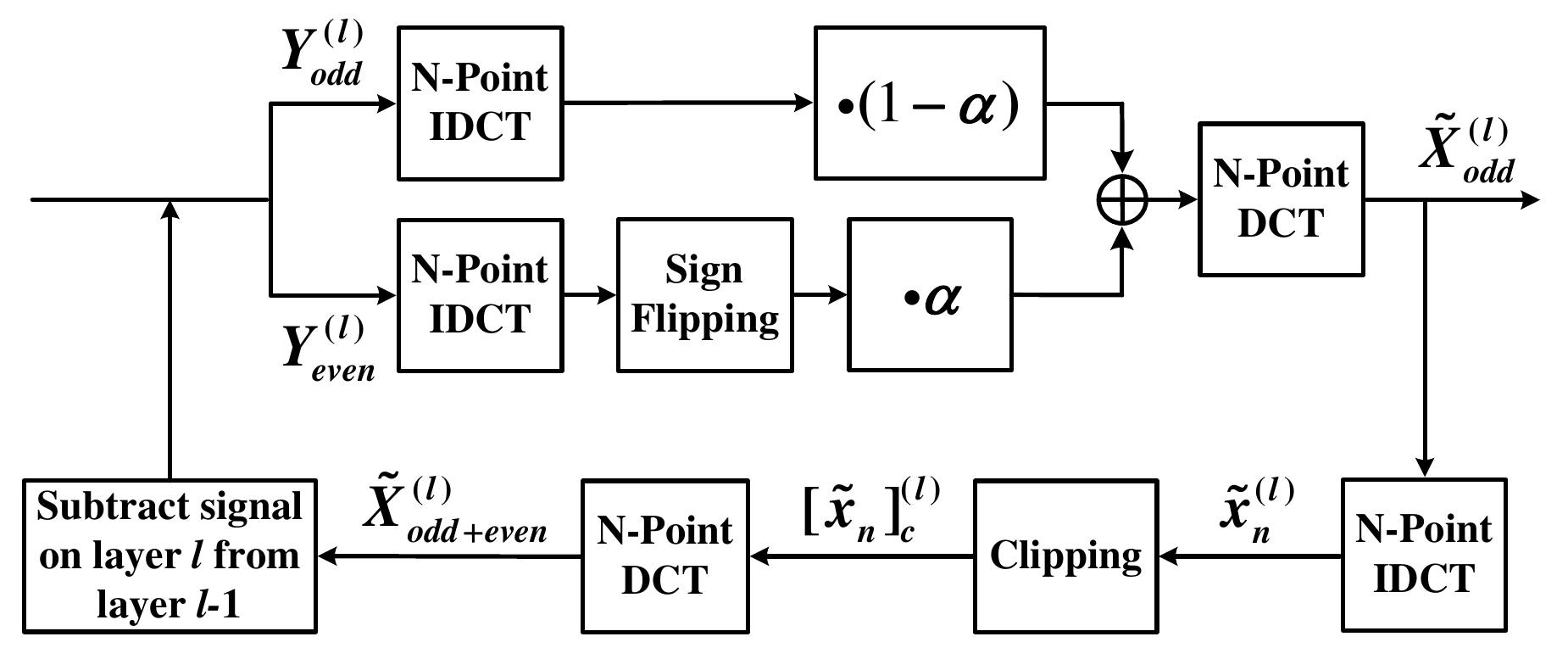}
\label{improved-receiver}}
\caption{(a) Conventional receiver structure of layered ACO-FOFDM; (b) Improved receiver structure of layered ACO-FOFDM}
\label{receiver structure}
\end{figure}

For traditional ACO-FOFDM system, the received signal ${X_{odd}}$ on the odd subcarrier can be represented as ${X_{odd}} = \frac{1}{2}DCT({x_n})$ and the received signal ${X_{even}}$ on the even subcarrier can be represented as ${X_{even}} = \\ \frac{1}{2}DCT(\left| {{x_n}} \right|)$, $x_n$ means the time domain signal on subcarrier $n$, $DCT(\cdot)$ means the DCT operation. These results are also suitable for higher layers in layered ACO-FOFDM system. The conventional receiver structure is shown in Fig. \ref{conventional-receiver}. Due to the asymmetric clipping operation of $l$-th layer brings clipping noise, which can influence the signal on higher layers, the first objective is to extract the signal $\tilde Y{_{odd}^{(1)}}$ on layer 1 from odd subcarriers, this operation can help us regenerate the clipping noise falling on even subcarriers. It is worth noting that the hard-decision \cite{SEE-OFDM} is applied to get the clipping noise, which means $\tilde Y{_{odd}^{(l)}}$ is the firm decision on the constellation values instead of the noisy constellation values with soft-decision. The hard-decision has better BER performance than soft-decision because more accurate constellation values are used to estimate the clipping noise. Then, the obtained signal on layer 1 can perform $N$-point IDCT, get the absolute value and perform $N$-point DCT, so that the clipping noise on even subcarriers of layer 1 is obtained. After subtracting the clipping noise generated by layer 1, the signal $\tilde Y{_{odd}^{(2)}}$ on layer 2 can be extracted. Afterwards, the demodulation process continues in a similar way for all the subsequent layers until the information at all layers is recovered.

\subsection{Improved Receiver Based on Diversity Combining}

The improved receiver is added after the conventional receiver. In the improved receiver, diversity combining technique is applied in each layer to improve the BER performance. From conventional receiver the signals on the odd subcarriers of all the layers are obtained, so that the diversity combining technique in improved receiver can be performed on the highest layer $L$ at first. Fig. \ref{improved-receiver} presents the improved receiver structure of layered ACO-FOFDM. If we take noise into consideration, the signals on odd and even subcarriers of layer $L$ can be represented as
\begin{equation}
Y_{odd}^{(L)}=X_{odd}^{(L)'}+N_{odd}^{(L)}=\frac{1}{2}DCT\left({x_{n}}^{(L)'}\right) + {N_{odd}^{(L)}}
\label{eq1}
\end{equation}
\begin{equation}
Y_{even}^{(L)}=X_{even}^{(L)'}+N_{even}^{(L)}=\frac{1}{2}DCT\left(\left| {{x_{n}}^{(L)'}} \right|\right) + {N_{even}^{(L)}}
\label{eq2}
\end{equation}
where $N_{odd}^{(L)}$ and $N_{even}^{(L)}$ are the frequency domain noises on odd and even subcarriers of layer $L$, respectively. It should be noted that $Y_{odd}^{(L)}$ and $Y_{even}^{(L)}$ are just the received signals on odd and even subcarriers, hard-decision is only needed to get the clipping noise.

The diversity combining technique can be applied to make use of the signal on both odd and even subcarriers. If we separate the signal on odd and even subcarriers, load $Y_{odd}^{(L)}$ and $Y_{even}^{(L)}$ to only the odd and even subcarriers of two separate IDCTs, we can get the results that
\begin{equation}
y_{n,odd}^{(L)} = IDCT(Y_{odd}^{(L)}) = \frac{1}{2}{{x_{n}}^{(L)'}} + {n_{n,odd}^{(L)}}
\label{eq3}
\end{equation}
\begin{equation}
\left|y_{n,even}^{(L)}\right| = IDCT(Y_{even}^{(L)}) = \frac{1}{2}{\left| {x_{n}}^{(L)'}\right|} + {n_{n,even}^{(L)}}
\label{eq4}
\end{equation}
where $n_{n,odd}^{(L)}$ and $n_{n,even}^{(L)}$ represent the time domain noises through the corresponding IDCT.

If we extract the polarity information from odd subcarriers to indicate the sign flipping of even subcarriers, we can obtain another useful signal from even subcarriers,
\begin{equation}
y_{n,even,f}^{(L)} = \left\{ \begin{array}{l}
{\kern 8pt} \left| {y_{n,even}^{(L)}} \right|{\kern 3pt},{\kern 10pt} {y_{n,odd}^{(L)}} > 0\\
- \left| {y_{n,even}^{(L)}} \right|{\kern 3pt},{\kern 10pt} {y_{n,odd}^{(L)}} \le 0
\end{array} \right.
\label{eq5}
\end{equation}

Then, the maximal-ratio combining technology can be employed as
\begin{equation}
{y_n}^{(L)} = (1-\alpha){y_{n,odd}^{(L)}}+\alpha{y_{n,even,f}^{(L)}}
\label{eq6}
\end{equation}
The combining coefficient $\alpha$ is related to signal to noise ratio (SNR) of $y_{n,odd}^{(L)}$ and $y_{n,even,f}^{(L)}$,
\begin{equation}
\alpha  = \frac{{SNR_{even}}}{{SNR_{odd}+ SNR_{even}}}
\label{eq7}
\end{equation}
where $SNR_{even}$ is the SNR of $y_{n,even,f}^{(L)}$ and $SNR_{odd}$ is the SNR of $y_{n,odd}^{(L)}$. The value of $\alpha$ is always a little bit less than 0.5, the reason is that in the sign flipping process, $y_{n,odd}^{(L)}$ is influenced by noise, there is a small chance that noise has opposite sign and larger amplitude compared with the corresponding signal, which can lead to sign flipping error in $y_{n,even,f}^{(L)}$. Hence $y_{n,even,f}^{(L)}$ always has less SNR value than $y_{n,odd}^{(L)}$. Finally, ${y_n}^{(L)}$ inputs DCT and the required signal is on the odd subcarriers of layer $L$.

Afterwards, we should perform diversity combining on layer $L-1$, but signal of layer $L$ falls on the even subcarriers of layer $L-1$. The influence from layer $L$ on the even subcarriers of layer $L-1$ can be eliminated as shown in Fig. \ref{improved-receiver}. The hard-decision constellation value $\tilde X_{odd}^{(L)}$ inputs IDCT, after zero clipping and DCT, we can get all the signal on layer $L$. By subtracting it from even subcarriers of layer $L-1$, diversity combining technique can be conducted on layer $L-1$. Afterwards, the diversity combining process continues in a similar way for all the subsequent layers until the information at all layers is recovered.

\section{Simulation Results and Discussion}
\begin{figure}[!t]
\centering
\includegraphics[width=\linewidth]{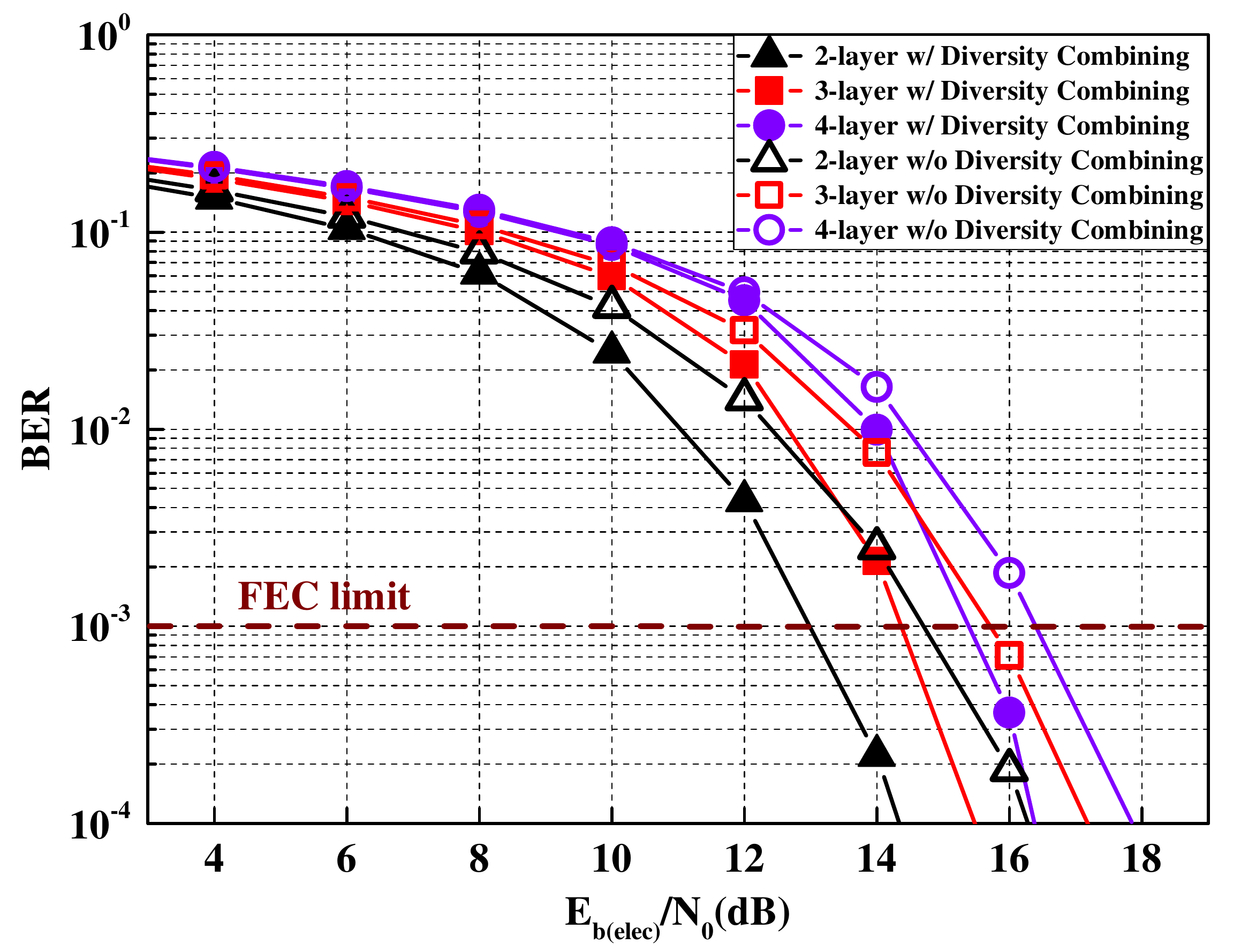}
\caption{BER performance comparison between improved receiver with diversity combining and conventional receiver without diversity combining in 4PAM layered ACO-FOFDM with different layers}
\label{4layer}
\end{figure}
The BER performance advantage of improved receiver is analyzed through simulation in additive white gaussian noise
(AWGN) channel. The DCT size is 256 in all the following simulations. Fig. \ref{4layer} reveals BER performance comparison between improved receiver with diversity combining and conventional receiver without diversity combining in layered ACO-FOFDM. The layer numbers are 2, 3 and 4 with 4PAM modulation on each layer. We can find with diversity combining the BER improvements are 1.76dB, 1.35dB and 1.02dB at forward error correction (FEC) limit (i.e., 1$\times10^{-3}$) in terms of $E_{b(elec)}/N_0$, $E_{b(elec)}/N_0$ denotes the ratio between electrical energy per bit and single-sided noise power spectral density. The BER performance advantages increase with the decreasing of FEC limit. Therefore, the improved receiver based on diversity combining can improve the BER performance of layered ACO-FOFDM system.

Afterwards, we make BER performance comparison among layered ACO-FOFDM system with or without improved receiver and DCO-FOFDM system at the same spectral efficiency. If the constellation sizes on different layers are the same, the spectral efficiency gap between layered ACO-FOFDM and DCO-FOFDM can never be eliminated because infinite number of layers are required. If arbitrary constellation size can be chosen on each layer, the spectral efficiency gap can be eliminated with only small number of layers. Taking practical implementation into consideration, large number of layers introduce more computational complexity, so we utilize 2-layer ACO-FOFDM system with proper selection of constellation sizes on different layers. The average signal power applied to each layer is proportion to the number of bits transmitted in each layer. The constellation size combination mode with the best BER performance for 2-layer ACO-FOFDM system is chosen according to Monte Carlo simulations, finding the optimal layer number and constellation size combination mode to get the best BER performance can be fulfilled in the future work.

\begin{figure}[!t]
\centering
\includegraphics[width=\linewidth]{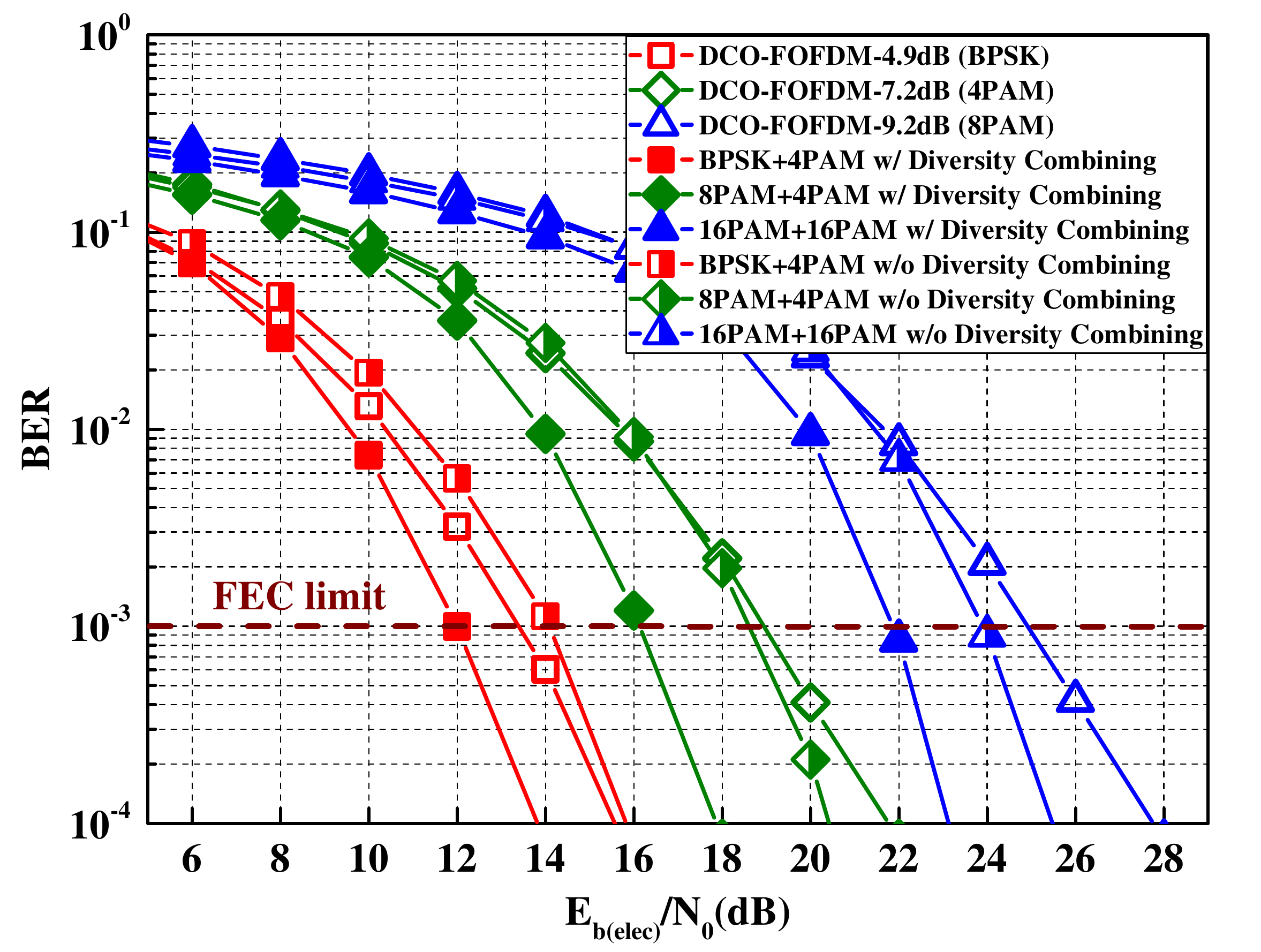}
\caption{BER as a function of $E_{b(elec)}/N_0$ for DCO-FOFDM system and layered ACO-FOFDM system with or without diversity combining. $E_{b(elec)}/N_0$ denotes the ratio between electrical energy per bit and single-sided noise power spectral density. The optimal DC-biases of DCO-FOFDM system are 4.9dB, 7.2dB and 9.2dB for BPSK, 4PAM and 8PAM as investigated in \cite{FOFDM}}
\label{elec}
\end{figure}
Figure \ref{elec} reveals the BER performance comparison among DCO-FOFDM system and 2-layer ACO-FOFDM system with or without diversity combining at different values of $E_{b(elec)}/N_0$. The spectral efficiencies are 1 bit/s/Hz (i.e., BPSK+4PAM), 2 bits/s/Hz (i.e., 8PAM+\\4PAM) and 3 bits/s/Hz (i.e., 16PAM+16PAM). For DCO-FOFDM system, the optimal DC-biases are set to get the best BER performance according to \cite{FOFDM}. From Fig. \ref{elec} we can find at FEC limit the BER performance of 2-layer ACO-FOFDM system without diversity combining is only a little bit better than DCO-FOFDM system when the spectral efficiency is 2 bits/s/Hz or 3 bits/s/Hz, and its BER performance is worse than DCO-FOFDM system when the spectral efficiency is 1 bit/s/Hz. However, at FEC limit, when the spectral efficiencies are 1 bit/s/Hz, 2 bits/s/Hz and 3 bits/s/Hz, the 2-layer ACO-FOFDM system with diversity combining can achieve 1.37dB, 2.89dB and 3.01dB BER gains compared with DCO-FOFDM system, and the improved receiver with diversity combining obtains 2.08\\dB, 2.54dB and 2.01dB BER gains compared with the system without diversity combining.

The BER performance against $E_{b(elec)}/N_0$ depends on the ratio between electrical energy per bit and single-sided noise power spectral density, but the main system constraint is usually the average transmitted optical power \cite{DC-biased}. So we also investigate the BER performance comparison among DCO-FOFDM system and 2-layer ACO-FOFDM system with or without diversity combining at different values of $E_{b(opt)}/N_0$. The conversion from $E_{b(elec)}/N_0$ to $E_{b(opt)}/N_0$ is demonstrated as following.

For DCO-FOFDM system, with unity optical power the relationship between $E_{b(elec)}/N_0$ and $E_{b(opt)}/N_0$ can be derived as \cite{DC-biased,Asymmetrically Clipped Optical Fast OFDM}
\begin{equation}
\frac{E_{b(opt)}}{N_0}= \frac{k^2}{1+k^2}\frac{E_{b(elec)}}{N_0}
\label{eq8}
\end{equation}
where $k$ is related to the DC-bias $B_{DC}$ that represented as $B_{DC}=10log_{10}(k^2+1)$ dB.

For 2-layer ACO-FOFDM system, the average electrical signal power and optical signal power can be expressed as \cite{Augmenting,Asymmetrically Clipped Optical Fast OFDM},
\begin{align}
&P_{elec}=E\left[x_n^2\right]=E\left[\left(\sum\limits_{l=1}^{2}{x_n^{(l)}}\right)^2\right]\nonumber\\
&{\kern 21pt}=E\left[\left({x_n^{(1)}}\right)^2\right]+E\left[\left({x_n^{(2)}}\right)^2\right]+
2E\left[x_n^{(1)}\right]E\left[x_n^{(2)}\right]\nonumber\\
&{\kern 21pt}=\frac{1}{2}\left({\sigma^{(1)}}\right)^2+\frac{1}{2}\left({\sigma^{(2)}}\right)^2+2\frac{\sigma^{(1)}}{\sqrt{2\pi}}\frac{\sigma^{(2)}}{\sqrt{2\pi}}
\label{eq9}
\end{align}
\begin{align}
&P_{opt}=E\left[x_n\right]=E\left[\sum\limits_{l=1}^{2}{x_n^{(l)}}\right]=\frac{\sigma^{(1)}}{\sqrt{2\pi}}+\frac{\sigma^{(2)}}{\sqrt{2\pi}}
\label{eq10}
\end{align}
where $\left({\sigma^{(l)}}\right)^2$ ($l=1,2$) is the variance of bipolar signal on layer $l$. Therefore, the ratio between $\left({\sigma^{(1)}}\right)^2$ and $\left({\sigma^{(2)}}\right)^2$ is the same as the ratio between average signal power of layer $1$ and layer $2$. Due to the average signal power applied to each layer is proportion to the number of bits transmitted in each layer, and half number of subcarriers in layer $2$ carry useful signal compared with layer $1$, the ratios between $\left({\sigma^{(1)}}\right)^2$ and $\left({\sigma^{(2)}}\right)^2$ for the spectral efficiency of 1 bit/s/Hz (i.e., BPSK+4PAM), 2 bits/s/Hz (i.e., 8PAM+4PAM) and 3 bits/s/Hz (i.e., 16PAM+16PAM) are 1, 3 and 2, which can be represented by the alphabet $p$. So for the case of $P_{opt}=1$\cite{DC-biased,Comparison of ACO-OFDM}, the relationship between $E_{b(elec)}/N_0$ and $E_{b(opt)}/N_0$ can be derived as
\begin{equation}
\frac{E_{b(opt)}}{N_0}= \frac{(1+\sqrt{p})^2}{(1+p)\pi+2\sqrt{p}}\frac{E_{b(elec)}}{N_0}
\label{eq11}
\end{equation}

\begin{figure}[!t]
\centering
\includegraphics[width=\linewidth]{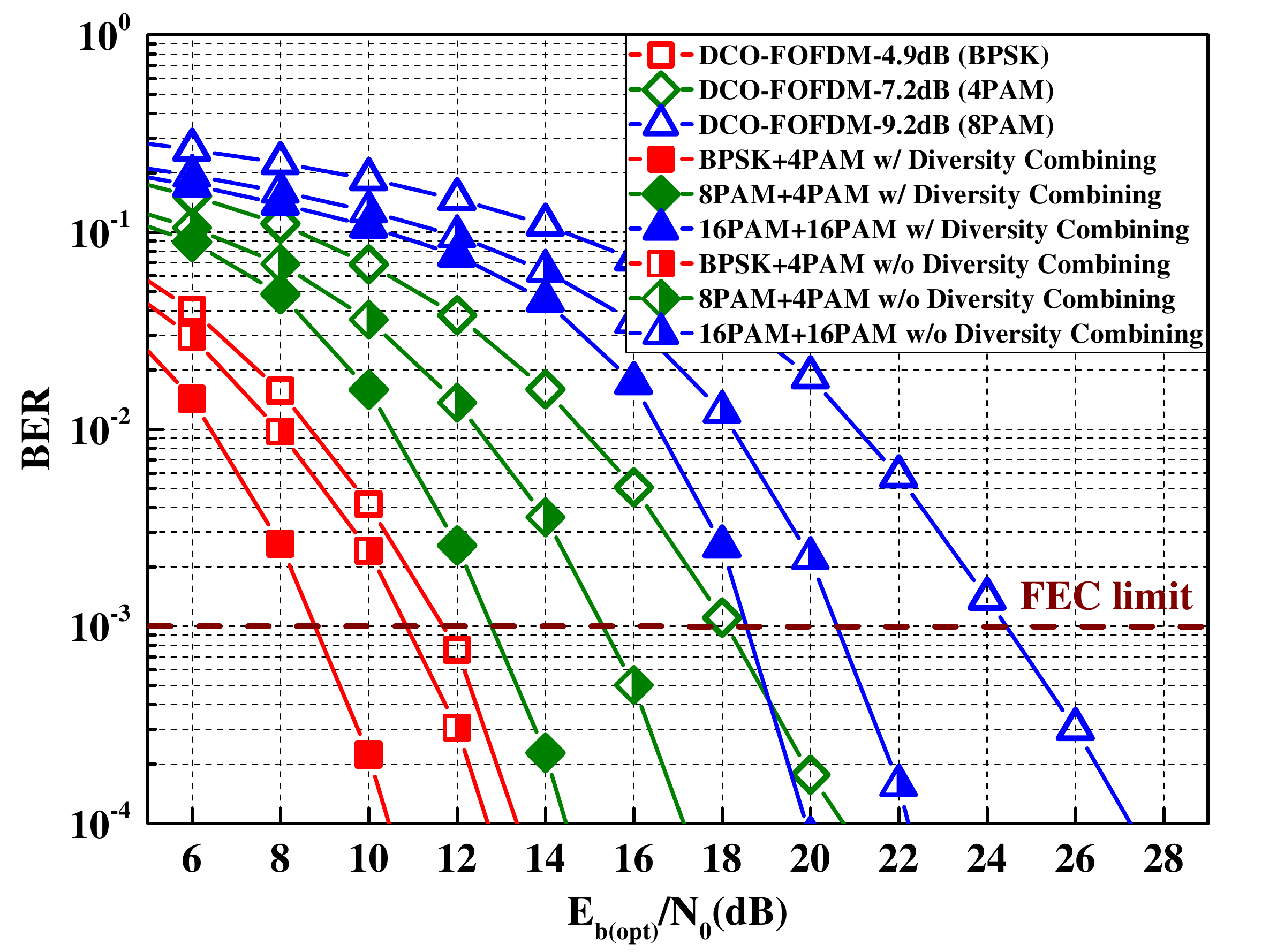}
\caption{BER as a function of $E_{b(opt)}/N_0$ for DCO-FOFDM system and layered ACO-FOFDM system with or without diversity combining. $E_{b(opt)}/N_0$ denotes the ratio between optical energy per bit and single-sided noise power spectral density. The optimal DC-biases of DCO-FOFDM system are 4.9dB, 7.2dB and 9.2dB for BPSK, 4PAM and 8PAM as investigated in \cite{FOFDM}}
\label{opt}
\end{figure}
The BER performance comparison among DCO-FOFDM system and 2-layer ACO-FOFDM system with or without diversity combining at different values of $E_{b(opt)}/N_0$ is depicted in Fig. \ref{opt}. The spectral efficiencies are 1 bit/s/Hz, 2 bits/s/Hz and 3 bits/s/Hz. The 2-layer ACO-FOFDM system without diversity combining is already able to achieve better BER performance than DCO-FOFDM system, with diversity combining more BER gains can be obtained. At FEC limit, the 2-layer ACO-FOFDM system with diversity combining can achieve 2.86dB, 5.26dB and 5.72dB BER gains compared with DCO-FOFDM system when the spectral efficiencies are 1 bit/s/Hz, 2 bits/s/Hz and 3 bits/s/Hz, and the system with diversity combining has the same BER gains as that in Fig. \ref{elec} compared with the system without diversity combining. Therefore, layered ACO-FOFDM system with improved receiver has low-cost property through the use of DCT, and is suitable for application in adaptive IM/DD systems with zero DC-bias.

\section{Conclusion}
In this paper, an improved receiver based on diversity combining in layered ACO-FOFDM system is proposed for IM/DD optical transmission systems. Layered ACO-FOFDM can compensate the weakness of traditional ACO-FOFDM in low spectral efficiency, the utilization of DCT instead of FFT can reduce the computational complexity without any influence on BER performance. At the transmitter, the superposition of multiple layers is performed in frequency domain to improve the spectral efficiency, the average signal power applied to each layer is proportion to the number of bits transmitted in each layer. At the receiver, conventional receiver obtains transmitted signal from lower layer to higher layer firstly, and improved receiver performs diversity combining technique from higher layer to lower layer to improve the BER performance. The BER performances of layered ACO-FOFDM system with improved receiver based on diversity combining and DC-offset FOFDM system with the optimal DC-bias are compared at the same spectral efficiency. Simulation results show that under different optical bit energy to noise power ratios, layered ACO-OFDM system with improved receiver has 2.86dB, 5.26dB and 5.72dB BER performance advantages at FEC limit over the DCO-FOFDM system when the spectral efficiencies are 1 bit/s/Hz, 2 bits/s/Hz and 3 bits/s/Hz, respectively. Layered ACO-FOFDM system with improved receiver based on diversity combining is suitable for application in the adaptive IM/DD systems with zero DC-bias.

\begin{acknowledgements}
This work was supported in part by National Natural Science Foundation of China (61427813, 61331010); National Key Research and Development Program (2016YFB0800302).
\end{acknowledgements}



\end{document}